\documentclass{aastex61}

\received{2018 February 18}
\revised{2018 April 3}
\accepted{2018 April 18}
\submitjournal{AAS Journals}

\shorttitle{On the detectability of Planet X with LSST}
\shortauthors{Trilling et al.}

\begin{document}

\title{On the detectability of Planet X with LSST}

\correspondingauthor{David E. Trilling}
\email{david.trilling@nau.edu}

\author{David E. Trilling}
\affil{Department of Physics and Astronomy, Northern Arizona
  University,
Flagstaff, AZ 86011, USA}
\affil{Lowell Observatory, Flagstaff, AZ 86001, AZ}

\author{Eric C. Bellm}
\affiliation{Department of Astronomy, University of Washington, Seattle, WA
98195, USA}

\author{Renu Malhotra}
\affiliation{Lunar and Planetary Laboratory, The University of
  Arizona, Tucson AZ 85721, USA}



\begin{abstract}
Two planetary mass objects 
in the far outer Solar System ---
collectively referred to here
as Planet X --- 
have recently been hypothesized to explain 
the orbital distribution of distant
Kuiper Belt Objects.
Neither planet is thought to be exceptionally
faint, but 
the sky
locations of these putative planets
are poorly constrained. Therefore,
a wide area survey is needed to detect 
these possible planets.
The Large Synoptic Survey Telescope
(LSST) will carry out an unbiased, large area
(around 18,000~deg$^2$), deep (limiting magnitude
of individual frames of~24.5) survey (the ``wide-fast-deep''
survey) of the
southern sky beginning in 2022, and is therefore an important
tool to search for these hypothesized planets.
Here we explore the effectiveness of LSST
as a search platform for these possible planets.
Assuming the current baseline cadence (which includes
the wide-fast-deep survey plus additional coverage)
we estimate that LSST will confidently detect or rule out the existence of Planet~X in 61\% of the entire sky.
At orbital distances up to $\sim$~75~au, Planet X could simply be found in the normal nightly moving object processing; at larger distances, it will require custom data processing.
We also discuss the implications
of a non-detection of Planet~X in LSST data.
\end{abstract}

\keywords{Kuiper Belt: general --- methods: observational ---
parallaxes --- planets and satellites: detection ---
surveys}



\section{Introduction} \label{intro}

The possibility of undiscovered planets in the solar system has
long fascinated astronomers and the public alike.
Recent studies of the orbital properties of very distant Kuiper belt
objects (KBOs) have identified several anomalies that may be
due to the gravitational influence
of one or more undiscovered planetary mass objects orbiting the Sun at distances
comparable to the distant KBOs.
\cite{trujillo2014} and \cite{sheppard2016} noted a clustering of the argument of perihelion (the angular position
of the perihelion relative to the ascending node of an orbit on the J2000 reference plane) of KBOs 
whose semi-major axes exceed 150~au.
Subsequently, \cite{batygin2016} and \citet{brown2016} noted a clustering of the longitudes of perihelion and of the orbit poles of the same group of distant KBOs. 
 \cite{malhotra2016} noted that the most distant KBOs have near-integer period ratios, suggestive of dynamical resonances with a massive perturber.

These orbital distribution peculiarities could be caused by an unseen massive body. 
\cite{trujillo2014} estimate a super-Earth mass object orbiting at a heliocentric distance $\gtrsim250$~au
(while noting that a range of parameters for an unseen parameter could
produce the observational signature that they observe);
\citet{brown2016} estimate a planet of mass 5--20~$M_\oplus$ in an orbit of semi-major axis 380--980~au, perihelion distance 150--350~au and moderately inclined ($\sim30^\circ$) to the ecliptic; and \cite{malhotra2016} suggest a $\sim10~M_\oplus$ planet in an orbit of semi-major axis $\sim665$~au of moderate eccentricity and two possible inclinations ($i\approx18^\circ$ or $i\approx48^\circ$ to the ecliptic).
The observational sample size for the above analyses is relatively small, 6--13~objects, depending upon choice of perihelion distance cut-off.
Finally, in
a separate line of argument,
\citet{bailey} and \citet{gomes} show that
the obliquity of the Sun of around 6~degrees can be
explained by a 10--20~M$_\oplus$ perturber with
a semi-major axis of 400--600~au.

Separately, for a larger sample of 
$\sim$160~distant KBOs whose semi-major axes are in the range 50--80~au,
\cite{volk} reported a strong deviation of the mid-plane from the
Solar System's invariable plane. 
Based on this deviation,
they suggest the presence of a smaller planetary
mass object of mass 0.1--2.4~$M_\oplus$ at distance 60--100~au in an orbit inclined to the ecliptic by a few to a few tens of degrees.

The predicted locations (in the sky) and brightnesses
for these massive unseen objects --- here referred to collectively as 
``Planet~X'' --- are sufficiently unconstrained 
that large area sky surveys 
must be carried out. Such surveys require
moderately large telescopes --- a distant planet
may be as bright as V=15 \citep{volk} or as faint as
V=22--25 \citep{brown2016} --- and very large fields
of view, as the search regions are at least
hundreds of square degrees, and could easily be 
thousands.  

The Large Synoptic Survey Telescope (LSST),
when it comes online in~2022, will easily meet
these criteria, with its single exposure depth
of {\tt r}$\sim$24.5 and field of view of 9.6~deg$^2$.
Furthermore, and more powerfully,
some 85\% of the ten-year LSST program will
be used for the ``wide-fast-deep'' (WFD) survey in 
which a large portion of the available sky (roughly
-60 to 0~degrees declination)
will be 
observed around 100~times in each of 
six filters ({\tt ugrizy}), over ten years.
(We note that the galactic plane is not included
in the WFD coverage.)
The total sky coverage of this WFD survey
will be around 18,000~deg$^2$, or around
44\% of the entire sky.
In other words,
LSST will be an excellent facility to search for
Planet X in the southern sky.



Here we explore the possibility that LSST
can be used to detect Planet X, if it exists.

\section{Detecting Planet X with LSST}

In order for LSST to detect Planet X several
simple requirements must be met, as follows.
(1) The planet must be within the nominal
LSST sky coverage in the current
baseline observing cadence ({\texttt minion\_1016}) 
as simulated by the LSST Operations Simulator 
\citep{connolly,delgado,reuter}.
(2) The planet must be bright enough to be
detected by LSST. In a single exposure this
implies {\tt r}$\leq$24.5, but through stacking images,
deeper searches are possible.
(3) The planet must have an on-sky rate of motion that is
detectable with LSST frames.
The timescale on which a distant planet
can be detected depends on its rate of motion
and therefore its distance, 
and different
rates require
different
data analysis techniques.
(4) The LSST cadence must be commensurate
with detecting the planet's motion.

In the sections that follow we assume that the ability
to detect point sources within LSST's sky coverage
is perfect. In other words, 
we assume that 
every point source above
the 5$\sigma$~limit of {\tt r}$\sim$24.5 is detected
perfectly.
We ignore the effects of crowded fields.

\subsection{Coverage and cadence}

\begin{figure}
\includegraphics[angle=270,width=15cm]{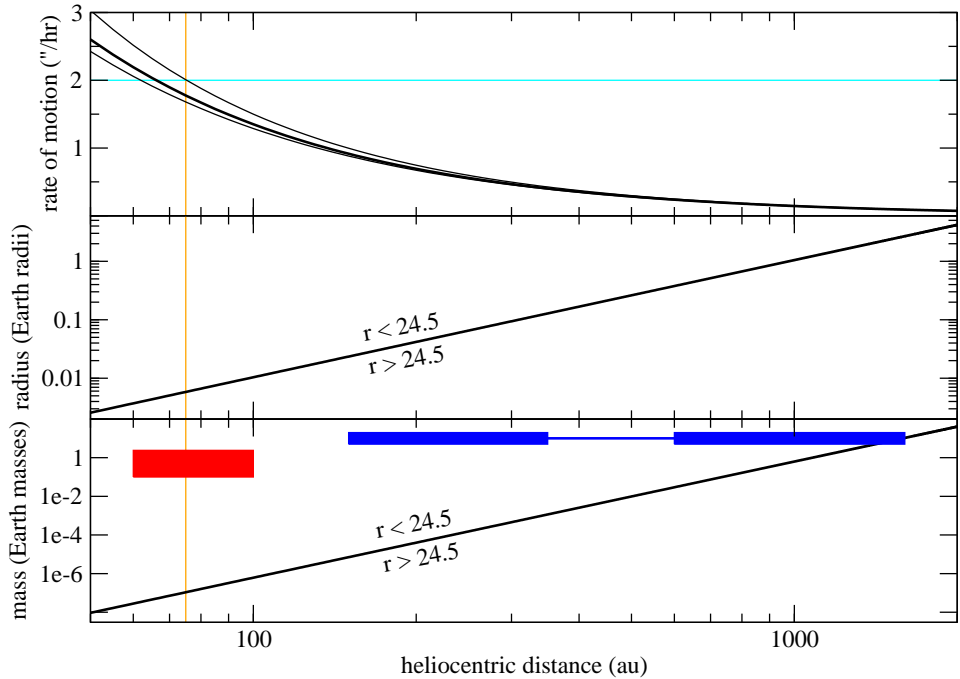}
\caption{{\em Top:} Sky rate of motion as 
a function of heliocentric distance.
The thick central line shows the rate for
an object on a circular orbit observed
at opposition.
The 
thin black lines show the
rates of motion for orbits with eccentricity
near unity for objects near
pericenter (lower curve)
and apocenter (upper curve).
For a perturber located at a particular heliocentric distance, the on-sky rate of motion for eccentric orbits lies either below or above the rate for circular orbits; the upper and lower curves show the range of on-sky rate of motion for the entire possible range of eccentricities, zero to unity.  
The light blue line at 2~arcsec/hr shows
the detectability limit for objects to be
detected in nightly processing; only
objects closer than 75~au (orange vertical
line in all panels) would be detected
in this standard processing.
{\em Middle and Bottom:} 
Radius (in Earth radii) and mass
(in Earth masses) of objects at the 
LSST detection limit of {\tt r}=24.5.
Detectable objects are above these lines.
In the bottom panel the Volk-Malhotra object
is shown as the red rectangle
 and 
the range of perihelion and aphelion
locations for the
Trujillo-Sheppard/Batygin-Brown object
is shown as the blue rectangles;
the thin blue line connects these regions
and shows the entire range of possible 
locations.
}
\label{threepanel}
\end{figure}

The LSST field of view is 
approximately circular with
a diameter of 3.5~deg \citep{ivezic}.
A distant Solar System object at 
\{60,1000\}~au 
has an apparent motion on the sky 
in the range of \{2.5,0.15\}~arcsec/hr 
or 
 \{6.1,0.37\} deg/year (Figure~\ref{threepanel}).
This rate of motion is dominated by the reflex motion of the Earth and is close to the parallactic motion; the orbital motion of a distant body contributes only a small fraction to its on-sky rate of motion.

Therefore, an object at 100~au will most
likely stay within one or two LSST
pointings per observing season, and 
an object at 1000~au will most likely
stay within one or two LSST pointings
over the ten year LSST baseline survey.
%
Here we consider the case of searching for
distant moving objects within a single observing season
(the months in a year when a given field is above
two airmasses at night and therefore could be observed
by LSST),
which means across multiple visits to a single
LSST pointing\footnote{To simplify the exposition, we describe searching for Planet X within a single LSST pointing.  However, LSST moving object searches are undertaken without respect for field boundaries.  Since adjacent fields have very similar observation cadences, Planet X would still be identified if it moved between two pointings in a single observing season.}.
(A pointing is a single
fixed piece of sky that LSST returns to over
and over again.)

For a pointing to be 
successfully searched we require five (or more)
visits within 45~days --- a nominal value
that approximates when a field is best placed
for observations ---
and that 
at least two pairs of observations are separated
by two (or more) days each.
The former requirement (45~days, five visits) ensures
that a reasonably long observational arc is obtained;
the latter (two day separation) ensures that 
object motion can be detected (though this turns out
to be a conservative assumption, as shown below).
We note that the nominal observing requirement 
for the LSST WFD survey is that any field that is observed once in a night
must be observed again $\sim$30~minutes later
in that same night.
Thus, there are enough
measurements across a season (45~days) 
to identify and measure the motion
of a distant object;
five
(or more) detections is a conservative requirement
that yields confidence in a 
consistent set of detections 
(minimizing the risk that
other astrophysical sources or noise have contaminated
the linked detections).
Requiring at least two
visits to be separated by more than two days
allows us the possibility of finding
distant objects, since the intra-night motion 
of objects at 1000~au is less than
one arcsecond and may not be easily
detected (as described below).

\begin{figure}
\plotone{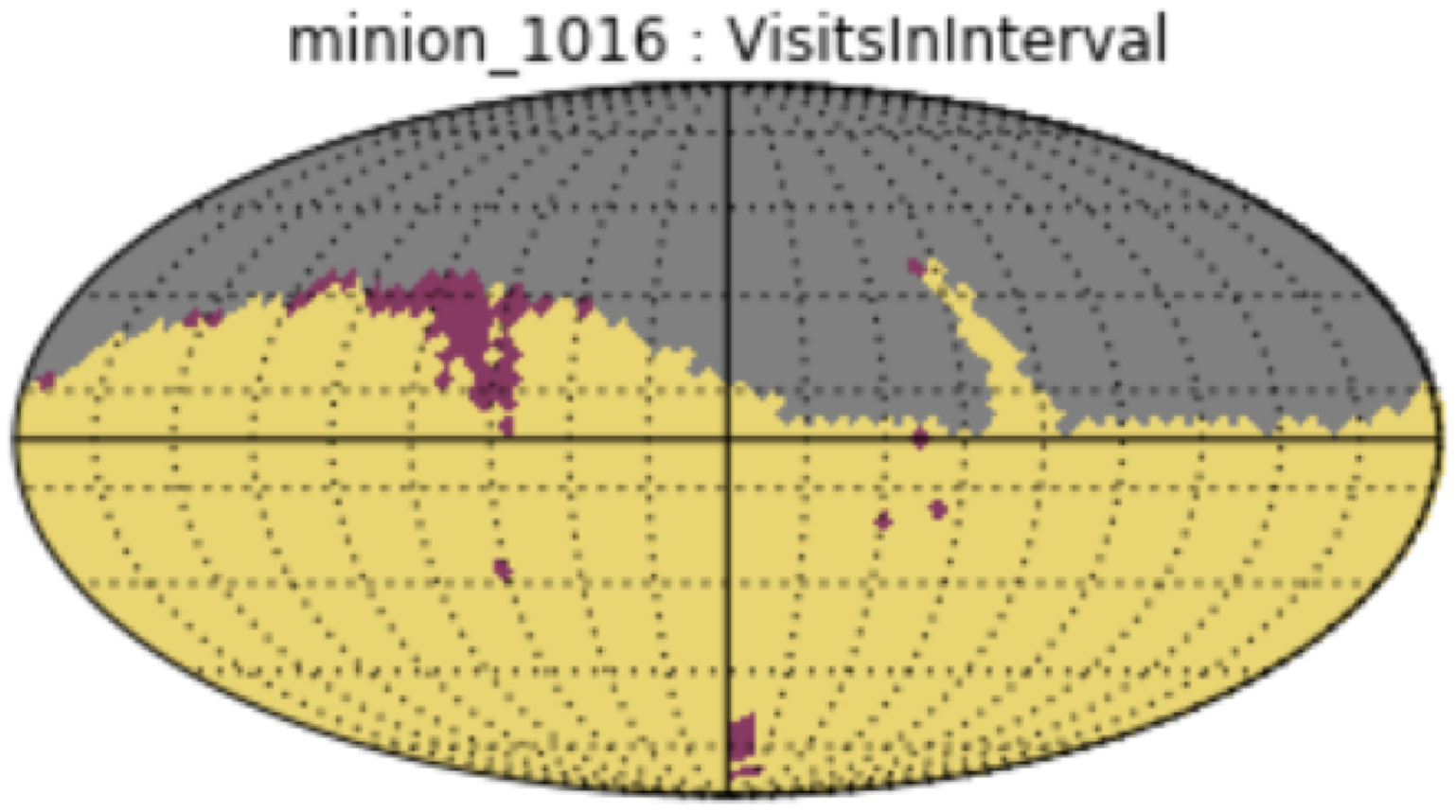}
\caption{Sky map of LSST fields,
in equatorial coordinates,
indicating detectability of outer solar
        system objects for the \texttt{minion\_1016} baseline cadence.
        Tan colored areas have five (or more)
visits within 45~days, with at least two pairs of observations separated
by two (or more) days each; 61\% of the entire sky
meets these criteria. Purple areas
are surveyed by LSST but
do not satisfy these criteria at
any point in the simulated ten-year survey.
(Grey regions
are not surveyed by
LSST.)
Of all the pointings in the current LSST baseline cadence,
96.9\% meet our cadence/visit requirement at least once during the ten
year survey.
In this figure grid lines are marked
every 20~degrees and the plot center is
RA of 180~degrees, with RA increasing to the
right from 0--360~degrees.
}
\label{coverage}
\end{figure}

Figure~\ref{coverage} shows the result of querying the current LSST
baseline cadence (\texttt{minion\_1016})
against these requirements. 
Tan colored areas show LSST pointings that satisfy
the above requirements, and purple show areas that
do not. (Grey regions are not surveyed by
LSST.)
We note that {\texttt{minion\_1016}} 
includes the WFD sky coverage as well as the North Ecliptic Spur, the South Celestial Cap, and the Galactic Plane. Some 44\% of the entire sky is surveyed 
in the baseline WFD survey, and 63\% of the sky is covered by some portion of the LSST survey. 
Of all the pointings in the current LSST baseline survey, 
96.9\% meet our requirements at least once during the ten year survey. 
This corresponds to 61\% of the entire sky. 

\subsection{Detections and data processing}

Almost all moving objects detected
in LSST data will be found through image differencing.
Standard LSST image differencing subtracts a pre-existing
coadded template image from 
each new visit image.
The current Moving Object Pipeline Software (MOPS)
design \citep[e.g.,][]{2016IAUS..318..282J, 2017AJ....154...13V}
first generates ``tracklets'' by
connecting difference image detections obtained in multiple visits within a
night.
The
median intra-night gap for visit pairs in the current LSST baseline cadence
(\texttt{minion\_1016} at the time of this writing) is about 30~minutes.  
Whether MOPS will recognize slow-moving outer Solar System objects
will depend on their rate of motion, the gap between visits within a night,
and the details of the LSST source association and MOPS pipelines.
Therefore,
given the expected\footnote{{\tt http://opsim.lsst.org/runs/reference\_run/minion\_1016/summary\_minion\_1016.pdf}}
LSST PSF of
around 0.8~arcsec, 
moving point sources
with separations of $\gtrsim$1~arcsec between
the two images
should be detected and linked through standard
LSST nightly processing. This corresponds to objects having a sky
motion $\gtrsim2$~arcsec/hr, equivalent to orbital distance $\lesssim$~75~au (Figure~\ref{threepanel}).
Therefore, given standard LSST nightly processing,
the parameter space of Planet X-like objects would be 
partially detectable for the Volk-Malhotra object but not at all detectable for the Trujillo-Sheppard/Batygin-Brown objects.

Possible planets that are beyond 75~au will
move too slowly to be detected in standard LSST
processing --- in this case, two detections within 
a night would be associated into a single non-moving
object ---
so additional strategies must be
employed to search for those objects.
If images from adjacent nights are compared,
objects with motions $\gtrsim$0.04~arcsec/hr
would be detectable (assuming a 24~hour baseline).
The corresponding distance for this motion
is $<$2000~au; all possible orbital parameters for
Planet X produce faster motions than this (Figure~\ref{threepanel}).
Here the process is the same as the nightly (standard
processing), but such night-to-night comparisons
and linking will not be provided by the LSST project.
(Night-to-night comparisons from non-adjacent nights
can be treated identically to those for adjacent nights.)

Direct imaging provides an alternative approach to identifying distant
moving objects.  Yearly LSST Data Release Processing will provide parallax
and proper motion estimates for objects in non-crowded fields;  
at {\tt r}$\sim$24, these have
accuracies of 2.9~mas  and 1.0~mas\,year$^{-1}$, respectively, over ten
years \citep{2009arXiv0912.0201L} --- easily sufficient
to detect the motion of all proposed Planet X-like objects. 


\section{Results and Discussion}

In Figure~\ref{threepanel}, the middle and lower panels show the radius (assuming
albedo of~0.5) and mass (assuming density
of 5.5~g/cm$^3$, same as the Earth's
density) of
objects at the photometry detection threshold of
LSST ({\tt r}=24.5). Detectable objects ({\tt r}$<$24.5) are
above the lines. \cite{brown2016}'s
proposed distant object solution --- 5--20~Earth mass object with perihelion of 150--350~au
and aphelion of 600--1600~au --- would be
easily detectable at perihelion,
and detectable at aphelion except for a
small slice of parameter space at the
largest possible distance
and smallest allowed mass (a region that 
becomes marginally more searchable through
stacking intra-night images).
In other words, for the 96.9\% of the 
LSST sky coverage where the cadence will
allow detection, the Trujillo-Sheppard/Batygin-Brown
object
is very likely to be detected.
Either night-to-night comparisons or direct imaging would be necessary
to detect this slow-moving object.
The Volk-Malhotra object is detected far more easily,
as it is much brighter (i.e., more massive)
than the detection limit shown in Figure~\ref{threepanel}.
In some cases, standard MOPS processing might reveal
the Volk-Malhotra object; in other cases, night-to-night
comparisons or direct imaging may be required.




In general, LSST should be able to detect
both
the Trujillo-Sheppard/Batygin-Brown object and 
the Volk-Malhotra object.
The Volk-Malhotra object would be closer and therefore
faster moving. (It is also predicted
to be brighter than the 
Trujillo-Sheppard/Batygin-Brown object.)
In general, greater sky rate of motion makes detection and linking with
MOPS easier.
Objects moving faster than about
0.7~arcsec/hr (i.e., closer than
about 200~au) should be readily detectable
in night-to-night processing, having moved some
17~arcsec in 24~hours. Such an object
will move only around 13~arcmin during the
45~day window specified above, and so would
be likely to stay within a single LSST pointing
during the entire observing season.



In these calculations we have used the current
LSST baseline cadence ({\tt minion\_1016}). The
LSST project is also actively considering a 
rolling cadence that increases the number of
visits to a given field in an observing season.
If anything,
the rolling cadence would increase the detectability
of Planet X, giving more detections in a single season
(since we have shown above that the most likely detection
scenario is with intra- or inter-night detections, not
year-to-year).

\citet{brown2016} aggregated a large number of previous and ongoing
surveys to show regions of the sky where their proposed object
can presently be ruled out and identified regions where it could still 
exist undetected. They find that the most likely
region on the sky for Planet~X is RA$\approx$2--10~hours
and declination of around -20~to~+10~degrees, 
where the object would be nearer its aphelion
and therefore slowly moving (roughly 0.1~arcsec/hr) and
faint (V of 22--25). 
This rate of motion would require night-to-night
comparisons, but this is not infeasible; the apparent
magnitude also is mostly brighter than the LSST detection
limit (Figure~\ref{threepanel}).
Interestingly, however, this region of sky is the 
part of the baseline LSST cadence that is most likely
to fail to meet our cadence requirements (Figure~\ref{coverage}).

LSST is likely to spend significant time surveying the
so-called North Ecliptic Spur: the part of the ecliptic plane
that is north of the nominal WFD survey coverage.
(Coverage of the North Ecliptic Spur is shown in
Figure~\ref{coverage} as the region north of the 
celestial equator.) The cadence for surveying this region
of the sky may be optimized for nearby Solar System objects,
and is not likely to significantly affect the discovery
space for distant Solar System objects. It is reasonable
to assume that the detection efficiency in the North
Ecliptic Spur is as good, or better,
than in the nominal WFD survey area.
Our probability of detecting Planet X with LSST
is thus most affected by the total sky area covered 
rather than the details of the cadence.

The high stellar density in the galactic plane
will make detections of distant moving 
objects there more difficult.
If Planet X exists there it is less likely 
to be detected than if it exists in less crowded
regions of the sky.
In the case of the non-detections of Planet X
the reduced
detection efficiency there
naturally weakens any conclusions
to be drawn from that region of sky,
but
those
data will still be useful to place constraints
once the detection efficiency is measured.

A non-detection with LSST would yield a map of regions of the sky where planets brighter than {\tt r}=24.5 
with a corresponding map of size and distance are ruled out;
this map would only apply to the 63\% of the sky that will be surveyed by LSST.
From such a map, we could compute the probability for ruling out the existence of a Mars-to-Earth size object
at 60--100~au and, 
similarly, for a Neptune-size object at 150--1600~au.
If these probabilities are close to unity, the recently proposed hypotheses for distant planets 
would lose support.
It would then be necessary to re-evaluate alternative explanations for the anomalies in the orbital properties of distant KBOs. Two alternatives are that the anomalies are statistical flukes, or that the anomalies indicate a relatively recent perturbation to the distant Solar System such as by a close stellar flyby.
The ``statistical fluke'' explanation is also testable by LSST because 
of the large number of distant KBOs that LSST will discover, which will allow much better statistical analysis of their orbital properties and re-assessment of the evidence for Planet~X.


\section{Conclusion}

If the putative Planet X --- either the Trujillo-Sheppard/Batygin-Brown
object or the Volk-Malhotra object --- exists, is within the
area of sky that LSST will cover, and is brighter 
than the LSST single exposure depth
then there is a very high probability that it will be 
detected, either with standard processing or through
custom processing.
For orbital distances closer than about 75 au, 
the rate of motion is fast enough
that it can be detected in the standard LSST
moving object nightly processing. More involved data 
processing is required to detect objects that are more
distant.
Should there be no detection
of distant planet(s) in LSST data, 
the orbits (and potential anomalies thereof) of
the many distant Kuiper
Belt Objects that LSST will discover
will help constrain the properties and
existence of Planet~X in regions of the 
sky not surveyed (or inadequately surveyed) by LSST.

As we were completing this work, we became aware of a third, independent line of argument by \citet{silsbee} who suggest that an unseen Mars-to-Earth sized body in an orbit inclined about 30 degrees to the ecliptic and of perihelion distance 40--70 au and semi-major axis less than 200 au could account for the properties of the detached KBOs (a population of KBOs with perihelion distance exceeding $\sim$38 au and semi-major axes 80--500 au).  This suggested object is similar in mass to the one suggested by Volk \& Malhotra (2017), spans a similar range of ecliptic latitude, but a larger range of heliocentric distance (40--360 au).  Our conclusions for the putative Planet X described above apply to this object as well.

\acknowledgements

DET acknowledges support from the Office of the Vice President
for Research at Northern Arizona University.
RM is grateful for research support from NASA (grant NNX14AG93G) and NSF (grant AST-1312498).
An anonoymous referee provided a number of suggestions
that improved this manuscript.

\vspace{5mm}

\software{LSST metrics analysis framework \citep{maf}}



\end{document}